\documentclass[11pt,a4paper]{article}
\RequirePackage{amsmath,amssymb}
\RequirePackage[dvipsnames,usenames]{color}
\usepackage{soul}
\usepackage{cite}
\usepackage{fullpage}
\usepackage[british]{babel}
\usepackage[latin1]{inputenc}
\usepackage[T1]{fontenc}
\usepackage[final]{showkeys} 
\usepackage[bookmarks]{hyperref}
\usepackage{amsthm}
\usepackage{graphicx}
\usepackage{subfigure}
\usepackage{braket}
\usepackage{mathrsfs}
\usepackage{bm}
\newcommand{\be}{\begin{equation}} \newcommand{\ee}{\end{equation}} 
 
\title{\bf A new approach to the thermodynamics \\ of scalar-tensor 
gravity}
\author{ Valerio~Faraoni$^{a}$\thanks{E-mail: vfaraoni@ubishops.ca}, $\ $ 
Andrea~Giusti$^{bc}$\thanks{E-mail: agiusti@phys.ethz.ch}, $\ $ and 
Andrea~Mentrelli$^{cd}$\thanks{E-mail: andrea.mentrelli@unibo.it}, \\ \\ 
$^a${\em Department of Physics \& Astronomy, Bishop's University} \\ {\em 
2600 College Street, Sherbrooke, Qu\'ebec, Canada J1M~1Z7} \\ \\ $^b${\em 
Institute for Theoretical Physics, ETH Zurich} \\ {\em 
Wolfgang-Pauli-Strasse 27, 8093 Zurich, Switzerland} \\ \\ $^c${\em Alma 
Mater Research Center on Applied Mathematics ($AM^2$),} \\ {\em University 
of Bologna, Bologna, Italy} \\ \\ $^d${\em Department of Mathematics, University of Bologna} \\ {\em Via Saragozza 8, 40123 Bologna, 
Italy} }
\begin{document}
\maketitle
\begin{abstract}
We discuss and expand a new approach to the thermodynamics 
of scalar-tensor gravity and its diffusion toward general relativity (seen 
as an equilibrium state) proposed in a previous Letter 
[Phys. Rev. D {\bf 103}, L121501 (2021)], 
upon which we build. We describe scalar-tensor gravity as an effective 
dissipative fluid and apply Eckart's first order thermodynamics to it, 
obtaining explicitly effective quantities such as heat flux, ``temperature 
of gravity'', viscosities, entropy density, plus an equation describing the 
``diffusion'' to Einstein gravity. These quantities, still missing in the 
usual thermodynamics of spacetime, are obtained with minimal assumptions.
Furthermore, we examine certain exact solutions of scalar-tensor gravity to test 
the proposed formalism and gain some physical insight on the 
``approach to equilibrium'' for this class of theories.
\par \null \par
\end{abstract}

\section{Introduction} \label{sec:1} \setcounter{equation}{0}

There seems to be a deep connection between thermodynamics and gravity, 
first observed in black hole thermodynamics, which produced an unexpected 
relation between the entropy and horizon area for stationary black holes 
and the temperatures of black hole and Rindler horizons. This connection 
took on a new meaning with Jacobson's seminal work \cite{Jacobson:1995ab} 
obtaining the Einstein field equation of general relativity (GR) as an 
equation of state, based only on thermodynamical considerations. This 
derivation of the Einstein equation and the ensuing ``thermodynamics of 
spacetime'' picture carry deep implications for gravity. Their main 
consequence would be that classical gravity is an emergent phenomenon 
instead of having a fundamental nature. If confirmed, this property would 
have radical consequences for quantum gravity as well. In the 
thermodynamics of spacetime picture, quantizing the Einstein equation would 
not be more meaningful than quantizing the macroscopic ideal gas equation 
of state, which cannot produce fundamental quantum results such as the 
energy spectrum and eigenfunctions of the hydrogen atom. In quantum 
gravity, entities such as the ``atoms of spacetime'' may not even exist or, 
if they do, they may have to be found using approaches radically different 
from the quantization of the Einstein equation.

A second idea,which is probably equally important, was proposed in 
Ref.~\cite{Eling:2006aw}, in which the authors derived the field equation 
of fourth order metric $f({\cal R}) $ gravity using only thermodynamics. 
This modification of GR, which contains an extra scalar degree of freedom 
$f'({\cal R})$ (see \cite{Sotiriou:2008rp, DeFelice:2010aj, Nojiri:2010wj} 
for reviews) would correspond to dissipative non-equilibrium 
``thermodynamics of gravitational theories'' in which a ``bulk viscosity of 
spacetime'' was introduced to explain dissipation \cite{Eling:2006aw}. By 
contrast, GR would correspond to a state of thermodynamic equilibrium 
\cite{Eling:2006aw}.

The works \cite{Jacobson:1995ab,Eling:2006aw} have generated a huge 
literature. In view of our new approach to this paradigm, it is useful to 
note that Ref.~\cite{Chirco:2010sw} has stressed the essential role of 
shear viscosity, while removing altogether bulk viscosity from the 
thermodynamical picture of $f({\cal R})$ gravity. In spite of the large 
literature, the equation(s) ruling the approach of modified gravity to the 
GR equilibrium state remain a mystery. Furthermore, the order parameter 
(presumably, the temperature) regulating this dissipative phenomenon has 
not yet been identified. Here we discuss in detail 
and expand a new approach proposed in 
Refs.~\cite{Faraoni:2021lfc,Giusti:2021sku} to the last two problems in 
the spirit 
of the thermodynamics of spacetime, but in a very different context. We 
consider the larger class of scalar-tensor theories of gravity 
\cite{BransDicke,ST-1, ST-2, ST-3} which contains $f({\cal R}) $ gravity as 
a subclass \cite{Sotiriou:2008rp, DeFelice:2010aj, Nojiri:2010wj}. 
Scalar-tensor gravity is a minimal modification of GR obtained by adding a 
massive scalar degree of freedom $\phi$ to the usual two massless spin two 
modes of GR contained in the metric tensor $g_{ab}$. The contribution of 
$\phi$ to the field equations can be described as an effective relativistic 
dissipative fluid \cite{Pimentel89,Faraoni:2018qdr}. Using this 
description, we apply Eckart's first order thermodynamics \cite{Eckart40} 
to this effective fluid and extract explicit expressions for the relevant 
effective thermodynamic quantities, including the heat current density, the 
``temperature of modified gravity'', the viscosity coefficients, and the 
entropy density.

To anticipate our findings: the product between the effective temperature 
${\cal T}$ and the thermal conductivity ${\cal K}$ is positive-definite and 
the GR equilibrium state corresponds to ${\cal K}{\cal T}=0$; the bulk 
viscosity vanishes, and the shear viscosity $\eta$ is negative. This 
unexpected sign could allow the entropy density $s$ to decrease, which is 
consistent with the fact that the $\phi$-fluid, seen as a thermodynamic 
system, is neither an isolated fluid~--~it exchanges energy with its 
``surroundings'' -- nor a real fluid. To proceed, we describe explicitly 
the approach of scalar-tensor gravity to the GR equilibrium state. In a 
sense, it is remarkable that, in spite of the well-known limitations of 
Eckart's first order thermodynamics, these explicit expressions and 
effective diffusion equation emerge from the formal identification of an 
effective fluid with a thermodynamic system, which would seem very unlikely 
{\em a priori}. The simplicity and minimality of assumptions of this new 
approach point again to some deeper connection between thermodynamics and 
gravity. Furthermore, in order to properly illustrate the physical 
implications of the proposed approach, we study the thermodynamical 
behavior of certain exact solutions of scalar-tensor gravity.

Let us review the basics of scalar-tensor gravity.  The scalar-tensor
 action in the Jordan frame is\footnote{We follow the notation of 
Ref.~\cite{Waldbook} and we use units in which Newton's constant $G$ and 
the speed of light $c$ are unity.} \be S_\text{ST} = \frac{1}{16\pi} \int 
d^4x \sqrt{-g} \left[ \phi {\cal R} -\frac{\omega(\phi )}{\phi} \, 
\nabla^c\phi \nabla_c\phi -V(\phi) \right] +S^\text{(m)} \,, 
\label{STaction} \ee where ${\cal R}$ is the Ricci scalar, the Brans-Dicke 
scalar $\phi>0$ is approximately the inverse of the effective gravitational 
coupling, $\omega(\phi)$ is the ``Brans-Dicke coupling'', $V(\phi)$ is a 
potential for the scalar field, and $S^\text{(m)}=\int d^4x \sqrt{-g} \, 
{\cal L}^\text{(m)} $ is the matter action.

By varying the action~(\ref{STaction}) with respect to the inverse metric 
$g^{ab}$ and to $\phi$, one obtains the (Jordan frame) field equations 
\cite{BransDicke, ST-1, ST-2, ST-3} \begin{eqnarray} G_{ab} \equiv {\cal 
R}_{ab} - \frac{1}{2}\, g_{ab} {\cal R} &=& \frac{8\pi}{\phi} \, 
T_{ab}^\text{(m)}
+ \frac{\omega}{\phi^2} \left( \nabla_a \phi \nabla_b \phi -\frac{1}{2} \, 
  g_{ab} \nabla_c \phi \nabla^c \phi \right) \nonumber\\
&&\nonumber\\
&\,& +\frac{1}{\phi} \left( \nabla_a \nabla_b \phi
- g_{ab} \Box \phi \right) -\frac{V}{2\phi}\, g_{ab} \,, \nonumber\\
&& \label{BDfe1} \\
\Box \phi &=& \frac{1}{2\omega+3} \left( \frac{8\pi T^\text{(m)} }{\phi} + 
\phi \, \frac{d V}{d\phi}
-2V -\frac{d\omega}{d\phi} \nabla^c \phi \nabla_c \phi \right) \,, 
 \nonumber\\
&& \label{BDfe2}
\end{eqnarray} where ${\cal R}_{ab}$ is the Ricci tensor and $ T^\text{(m)} 
\equiv g^{ab}T_{ab}^\text{(m)} $ is the trace of the matter stress-energy 
tensor $T_{ab}^\text{(m)} $.
\section{Effective scalar field fluid} \label{sec:2} 
\setcounter{equation}{0}

Here we summarize the formulas for the effective field fluid derived in 
\cite{Pimentel89,Faraoni:2018qdr} that are needed in the calculations of 
this paper.

\subsection{Kinematic quantities}

Let us begin with the kinematic quantities of the effective $\phi$-fluid 
\cite{Ellis71} (here we provide extra expressions for the kinematic 
quantities and for the effective fluid quantities which were not given in 
\cite{Faraoni:2021lfc}, but are handy for calculations). The $\phi$-fluid 
description is natural when the gradient $\nabla^a \phi$ is timelike and 
can be used to construct the effective fluid four-velocity \be u^a = 
\frac{\nabla^a \phi}{\sqrt{ -\nabla^e \phi \nabla_e \phi }} 
\label{4-velocity} \ee normalized to $u^c u_c=-1$. The $3+1$ splitting of 
spacetime into the time direction $u^c$ and the 3-dimensional space of the 
comoving observers of this effective fluid (with four-tangent $u^c$) 
follows. Their 3-space is endowed with the Riemannian metric \be h_{ab} 
\equiv g_{ab} + u_a u_b \ee and ${h_a}^b$ is the projection operator on 
this 3-space, then \begin{eqnarray} h_{ab} u^a &=& h_{ab}u^b=0 \,,\\
&&\nonumber\\
{h^a}_b \, {h^b}_c &=& {h^a}_c \,, \;\;\;\;\;\;  {h^a}_a=3 \,. 
\end{eqnarray} The effective fluid four-acceleration is $ \dot{u}^a \equiv 
u^b \nabla_b u^a $ and is orthogonal to the four-velocity, $\dot{u}^c 
u_c=0$ (exceptions to this rule, which include 
Friedmann-Lema\^itre-Robertson-Walker spaces, particles with variable mass, 
{\em etc.} \cite{Faraoni:2020ejh} will not be considered in this work).

The projection of the velocity gradient onto the 3-space of the comoving 
observers is the purely spatial tensor \be V_{ab} \equiv {h_a}^c \, {h_b}^d 
\, \nabla_d u_c \,. \label{Vab} \ee It splits into symmetric and 
antisymmetric parts, with the symmetric part further decomposed into 
trace-free part and pure trace as \be V_{ab}= \theta_{ab} +\omega_{ab} 
=\sigma_{ab} +\frac{\theta}{3} \, h_{ab}+ \omega_{ab} \,, \ee with 
$\theta_{ab}=V_{(ab)}$ the expansion tensor (symmetric part of $V_{ab}$) 
with trace $\theta\equiv {\theta^c}_c =\nabla^c u_c $; the vorticity tensor 
$\omega_{ab}=V_{[ab]}$ is its antisymmetric part, while the trace-free 
shear tensor is \be \sigma_{ab} \equiv \theta_{ab}-\frac{\theta}{3}\, 
h_{ab} \,. \ee $V_{ab}$, $\theta_{ab}$, $\sigma_{ab}$, and $\omega_{ab}$ 
are purely spatial tensors, \be \theta_{ab}u^a = \theta_{ab}u^b = 
\omega_{ab} \, u^a = \omega_{ab} \, u^b = \sigma_{ab}u^a = \sigma_{ab} u^b 
= 0 \,, \ee while ${\sigma^a}_a={\omega^a}_a=0$. The shear scalar $\sigma$ 
and vorticity\footnote{Here there is no risk of confusing the vorticity 
scalar with the Brans-Dicke coupling because the former is always zero as 
the effective $\phi$-fluid is irrotational.} scalar $\omega$ are 
\begin{eqnarray} \sigma^2 & \equiv &\frac{1}{2} \, \sigma_{ab}\sigma^{ab} 
\geq 0 \,,\\
&&\nonumber\\
\omega^2 & \equiv &\frac{1}{2} \, \omega_{ab}\omega^{ab} \geq 0\,. 
\end{eqnarray} The velocity gradient decomposes as \cite{Ellis71} \be 
\nabla_b u_a = \sigma_{ab}+\frac{\theta}{3} \, h_{ab} +\omega_{ab} - 
\dot{u}_a u_b =V_{ab} -\dot{u}_a u_b \,. \label{ecce} \ee Projecting 
Eq.~(\ref{ecce}) onto $u^c$ produces $\dot{u}_a$, while projecting it onto 
the 3-space orthogonal to $u^a$ yields $V_{ab}$.

When these general definitions \cite{Ellis71, Waldbook} are specialized to 
our effective $\phi$-fluid, one obtains \cite{Faraoni:2018qdr} \be h_{ab}= 
g_{ab}-\frac{ \nabla_a\phi \nabla_b \phi}{ \nabla^e\phi \nabla_e\phi} \,, 
\ee \be \nabla_b u_a = \frac{1}{ \sqrt{ -\nabla^e\phi \nabla_e \phi}} 
\left( \nabla_a \nabla_b \phi -\frac{ \nabla_a \phi \nabla^c \phi \nabla_b 
\nabla_c \phi}{\nabla^e\phi \nabla_e \phi} \right) \,. \ee The 
4-acceleration then reads\footnote{It is straightforward to check that 
$\dot{u}_c u^c=0$ using Eqs.~(\ref{acceleration}) and~(\ref{4-velocity}).} 
\be \dot{u}_a = u^c \nabla_c u_a = \left( -\nabla^e \phi \nabla_e \phi 
\right)^{-2} \nabla^b \phi \Big[ (-\nabla^e \phi \nabla_e \phi)  \nabla_a 
\nabla_b \phi + \nabla^c \phi \nabla_b \nabla_c \phi \nabla_a \phi \Big] 
\,. \label{acceleration} \ee

The (timelike) worldlines of the $\phi$-fluid are geodesics (equivalently, 
this fluid is a dust) if and only if \begin{equation} \nabla^e \phi 
\nabla_{[e} \phi \nabla_{a]} \nabla_b \phi \nabla^b \phi = 0 \,, 
\end{equation} from which it follows that \be \nabla^b\nabla^c \phi 
\nabla_b\nabla_c \phi = -\frac{ \nabla^a\phi\nabla^b \phi \nabla_a\nabla_b 
\phi}{ \nabla^e\phi\nabla_e \phi } \ee for a geodesic flow. In this case, 
$V_{ab}$ reduces to \begin{eqnarray} V_{ab} = \frac{ \nabla_a \nabla_b \phi 
}{ \left( -\nabla^e \phi \nabla_e \phi \right)^{1/2} } +\frac{ \left( 
\nabla_a \phi \nabla_b \nabla_c \phi + \nabla_b \phi \nabla_a \nabla_c \phi 
\right) \nabla^c \phi }{ \left( -\nabla^e \phi \nabla_e \phi \right)^{3/2} 
}
+ \frac{ \nabla_d \nabla_c \phi \nabla^c \phi \nabla^d \phi }{\left( 
  -\nabla^e \phi \nabla_e \phi \right)^{5/2} } \, \nabla_a \phi \nabla_b 
  \phi \,. \end{eqnarray} Since the $\phi$-fluid four-velocity $u^c$ is 
  derived from a scalar field gradient, this fluid is irrotational, 
  $\omega_{ab}=\omega^2=0$, leaving \be V_{ab} = \theta_{ab} \,, \;\;\;\;\; 
  \nabla_b u_a = \theta_{ab} - \dot{u}_a u_b \,. \ee Moreover, the vector 
  field $u^a$ is hypersurface-orthogonal and the line element becomes 
  diagonal in adapted coordinates. In other words, the existence of a 
  foliation of 3-dimensional hypersurfaces $\Sigma$ with Riemannian metric 
  $h_{ab}$ orthogonal to $u^a$ is guaranteed \cite{Ellis71, Waldbook}.

Since $u_a \dot{u}^a=0$, the expansion scalar~(\ref{ecce})  reduces to 
\begin{eqnarray} \theta = \nabla_a u^a = \frac{ \square \phi}{ \left 
(-\nabla^e \phi \nabla_e \phi \right)^{1/2} } + \frac{ \nabla_a \nabla_b 
\phi \nabla^a \phi \nabla^b \phi }{ \left( -\nabla^e \phi \nabla_e \phi 
\right)^{3/2} } \,, \label{thetaScalar} \end{eqnarray} while 
\begin{eqnarray}
 \sigma_{ab}
&=& \left( -\nabla^e \phi \nabla_e \phi \right)^{-3/2} \left[
-\left( \nabla^e \phi \nabla_e \phi \right) \nabla_a \nabla_b \phi
- \frac{1}{3} \left( \nabla_a \phi \nabla_b \phi - g_{ab} \, \nabla^c \phi 
  \nabla_c \phi \right) \square \phi \right.\nonumber\\
     &&\nonumber\\
     &\, & \left. - \frac{1}{3} \left( g_{ab} + \frac{ 2 \nabla_a \phi
\nabla_b \phi }{ \nabla^e \phi \nabla_e \phi }
 \right) \nabla_c \nabla_d \phi \nabla^d \phi \nabla^c \phi + \left( 
\nabla_a \phi \nabla_c \nabla_b \phi + \nabla_b \phi \nabla_c \nabla_a \phi 
\right) \nabla^c \phi \right] \,, \label{sheartensor}\\
&&\nonumber\\
\sigma & \equiv & \left( \frac{1}{2} \, \sigma^{ab} \sigma_{ab} 
\right)^{1/2}
 = ( -\nabla^e \phi \nabla_e \phi)^{-3/2} \left\{ \frac{1}{2} \left( 
\nabla^e \phi \nabla_e \phi \right)^2 \left[ \nabla^a \nabla^b \phi 
\nabla_a \nabla_b \phi - \frac{1}{3} \left( \square \phi \right)^2 \right] 
\right. \nonumber\\
 &&\nonumber\\
&\, & \left.  + \frac{1}{3} \left( \nabla_a \nabla_b \phi \nabla^a \phi
\nabla^b \phi \right)^2 - \left( \nabla^e \phi \nabla_e \phi \right) \left( 
\nabla_a \nabla_b \phi \nabla^b \nabla_c \phi - \frac{1}{3} \, \square \phi 
\nabla_a \nabla_c \phi \right) \nabla^a \phi \nabla^c \phi \right\}^{1/2} 
\,. \nonumber\\
&&\label{sigmaScalar}
\end{eqnarray}
\subsection{Effective stress-energy tensor of the $\phi$-fluid}
The effective stress-energy tensor of the Brans-Dicke-like field that one 
reads off the right hand side of Eq.~(\ref{BDfe1}) is 
\begin{eqnarray} 
8\pi T_{ab}^{(\phi)} &=& \frac{\omega}{\phi^2} \left( \nabla_a \phi 
\nabla_b \phi -
 \frac{1}{2} \, g_{ab} \nabla^c \phi \nabla_c \phi \right)  + 
 \frac{1}{\phi} \left( \nabla_a \nabla_b \phi -g_{ab} \square \phi \right)
- \frac{V}{2 \phi} \, g_{ab} \,. \label{BDemt} 
\end{eqnarray} 
$T_{ab}^{(\phi)}$ takes the form of an imperfect fluid energy-momentum  
tensor \cite{Pimentel89,Faraoni:2018qdr} 
\be 
T_{ab} = \rho u_a u_b + q_a u_b + q_b u_a + \Pi_{ab} 
\,,\label{imperfectTab} 
\ee 
where the effective energy density, heat flux density, stress tensor, 
isotropic pressure, and anisotropic stresses (the trace-free part 
$\pi_{ab}$ of the stress tensor $\Pi_{ab}$) in the comoving frame are 
\begin{eqnarray} 
\rho &=& T_{ab} 
  u^a u^b \,, \label{rhophi}\\ q_a & =& -T_{cd} \, u^c {h_a}^d \,, 
  \label{qphi}\\
 \Pi_{ab} &= & Ph_{ab} + \pi_{ab} = T_{cd} \, {h_a}^c \, {h_b}^d \,, 
\label{Piphi}\\
    P &=& \frac{1}{3}\, g^{ab}\Pi_{ab} =\frac{1}{3} \, h^{ab} T_{ab} \,, 
\label{Pphi}\\
    \pi_{ab} &=& \Pi_{ab} - Ph_{ab} \,, \label{piphi} 
\end{eqnarray} 
respectively. Here we have set the bulk viscous pressure to zero in the 
most economical interpretation of the effective $\phi$-fluid (we refer the 
reader to \cite{Faraoni:2018qdr,Faraoni:2021lfc} for details). The heat 
flux density is purely spatial, 
\be 
q_c u^c = 0 \ee and \be \Pi_{ab} 
u^b=\pi_{ab} u^b = \Pi_{ab} u^a=\pi_{ab} u^a = 0 \, , \,\,\,\,\,\,\, 
{\pi^a}_a = 0 \,. 
\ee 
The explicit expressions of the effective 
stress-energy quantities are \cite{Pimentel89,Faraoni:2018qdr} 
\begin{eqnarray} 
8 \pi \rho^{(\phi)} &=& -\frac{\omega}{2\phi^2} \, 
\nabla^e \phi \nabla_e \phi + \frac{V}{2\phi} + \frac{1}{\phi} \left( 
\square \phi - \frac{ \nabla^a \phi \nabla^b \phi \nabla_a \nabla_b \phi}{ 
\nabla^e \phi \nabla_e \phi } \right)  \,,\label{effdensity}\\
&&\nonumber\\
\nonumber 8 \pi q_a^{(\phi)} &=& \frac{\nabla^c \phi \nabla^d \phi}{\phi
  \left(-\nabla^e \phi \nabla_e \phi \right)^{3/2} } \, \Big( \nabla_d \phi 
\nabla_c \nabla_a \phi
- \nabla_a \phi \nabla_c \nabla_d \phi \Big) \\ &&\nonumber\\ &=& 
  -\frac{\nabla^c \phi \nabla_a \nabla_c\phi}{\phi \left( -\nabla^e \phi
\nabla_e\phi \right)^{1/2}} - \frac{ \nabla^c\phi \nabla^d\phi 
\nabla_c\nabla_d \phi}{\phi \left( -\nabla^e\phi \nabla_e\phi \right)^{3/2} 
} \, \nabla_a \phi \,, \label{eq:q} \\ &&\nonumber\\ 8 \pi 
  \Pi_{ab}^{(\phi)} &=& (-\nabla^e \phi \nabla_e \phi)^{-1} \left[ \left(
- \frac{\omega}{2\phi^2}\, \nabla^e \phi \nabla_e \phi - \frac{\square 
  \phi}{\phi} - \frac{V}{2 \phi} \right) \Big( \nabla_a \phi \nabla_b \phi 
  - g_{ab} \nabla^e \phi \nabla_e \phi \Big) \right.  \nonumber\\ 
    &&\nonumber\\
&\, & \nonumber \left.  - \frac{\nabla^d
\phi}{\phi} \left( \nabla_d \phi \nabla_a \nabla_b \phi - \nabla_b \phi 
\nabla_a \nabla_d \phi - \nabla_a \phi \nabla_d \nabla_b \phi + \frac{ 
\nabla_a \phi \nabla_b \phi \nabla^c \phi \nabla_c \nabla_d \phi}{ \nabla^e 
\phi \nabla_e \phi } \right) \right] \\
&&\nonumber\\
&=& \left( -\frac{\omega}{2\phi^2} \, \nabla^c \phi \nabla_c \phi
-\frac{\Box\phi}{\phi} -\frac{V}{2\phi} \right) h_{ab} +\frac{1}{\phi} \, 
{h_a}^c {h_b}^d \nabla_c \nabla_d \phi \,, \label{eq:effPi2}\\
&&\nonumber\\
8 \pi P^{(\phi)} & = & - \frac{\omega}{2\phi^2} \, \nabla^e \phi \nabla_e 
\phi - \frac{V}{2\phi} - \frac{1}{3\phi} \left( 2\square \phi + 
\frac{\nabla^a \phi \nabla^b \phi \nabla_b \nabla_a \phi }{\nabla^e \phi 
\nabla_e \phi } \right) \,, \label{effpressure}\\
&&\nonumber\\
8 \pi \pi_{ab}^{(\phi)} &=& \frac{1}{\phi \nabla^e \phi \nabla_e \phi } 
\left[ \frac{1}{3} \left( \nabla_a \phi \nabla_b \phi - g_{ab} \nabla^c 
\phi \nabla_c \phi \right) \left( \square \phi - \frac{ \nabla^c \phi 
\nabla^d \phi \nabla_d \nabla_c \phi }{ \nabla^e \phi \nabla_e \phi } 
\right) \right. \nonumber\\
&&\nonumber\\
&\, & \left. + \nabla^d \phi \left( \nabla_d \phi \nabla_a \nabla_b
\phi - \nabla_b \phi \nabla_a \nabla_d \phi - \nabla_a \phi \nabla_d 
\nabla_b \phi + \frac{ \nabla_a \phi \nabla_b \phi \nabla^c \phi \nabla_c 
\nabla_d \phi }{ \nabla^e \phi \nabla_e \phi } \right) \right] \,, 
\label{piab-phi} \nonumber \\
& &
\end{eqnarray} 
and 
\be 
8 \pi T^{(\phi)} \equiv 8 \pi g^{ab}T_{ab}^{(\phi)}
 = - \frac{\omega}{ \phi^2} \, \nabla^c \phi \nabla_c \phi - \frac{3\square 
\phi}{\phi} - \frac{2V}{\phi} \,.\label{eq:efftrace} 
\ee

Alternative expressions for $\rho^{(\phi)} $ and $P^{(\phi)} $ are obtained 
by replacing $\square \phi$ with the help of Eq.~(\ref{BDfe2}): 
\begin{eqnarray} 
8 \pi \rho^{(\phi)} &\!\!=\!\!& - \frac{\omega}{2\phi^2} 
\, \nabla^e \phi \nabla_e \phi + \frac{V}{2\phi} \left( \frac{2\omega -1}{2 
\omega + 3} \right) \nonumber\\
&&\nonumber\\
&\, & +
\frac{1}{\phi} \left[ \frac{1}{2\omega + 3} \left( \phi \, \frac{dV}{d\phi} 
- \nabla^e \phi \nabla_e \phi \, \frac{d\omega}{d\phi} \right) - \frac{ 
\nabla^a \phi \nabla^b \phi \nabla_a \nabla_b \phi}{ \nabla^e \phi \nabla_e 
\phi} \right] \,, \label{rhophi2}\\
&&\nonumber\\
8 \pi P^{(\phi)} &\!\!=\!\!& - \frac{\omega}{2\phi^2} \, \nabla^e \phi 
\nabla_e \phi - \frac{V}{6\phi} \, \frac{ \left(6\omega + 
1\right)}{\left(2\omega
+  3\right)} \nonumber\\ &&\nonumber\\ &\, & - \frac{1}{3\phi} \left[ 
  \frac{2}{2\omega+3} \Big( \phi \, \frac{dV}{d\phi} - \nabla^e \phi 
  \nabla_e \phi \, \frac{d\omega}{d\phi} \Big) + \frac{
\nabla^a \phi \nabla^b \phi \nabla_b \nabla_a \phi }{\nabla^e \phi \nabla_e 
\phi } \right] \,. \label{Pphi2} 
\end{eqnarray}

In general, the effective fluid stress-energy tensor $T_{ab}^{(\phi)} $ 
does not satisfy any energy condition because it contains second 
derivatives of $\phi$ (which can have either sign) together with the more 
standard squares of first derivatives (which are instead 
positive-definite), preventing conclusions. For reference, we list the 
energy conditions, although they will be violated in general by the 
$\phi$-fluid\footnote{See Refs.~\cite{Kolassisetal88, Pimenteletal6} for a 
discussion of the energy conditions for an imperfect fluid with respect to 
arbitrary (timelike) observers.} 
\cite{Faraoni:2018qdr,Barcelo:2000zf,Bellucci:2001cc}.  The weak energy 
condition ($T_{ab} t^a t^b \geq 0$ for all timelike vectors $t^a$ 
\cite{Waldbook}) becomes 
\begin{eqnarray} 
T_{ab}^{(\phi)} u^a u^b = 
-\frac{\omega}{2\phi} \, \nabla^e \phi \nabla_e \phi + \frac{V}{2} + 
\square \phi - \frac{ \nabla^a \phi \nabla^b \phi \nabla_a \nabla_b \phi 
}{\nabla^e \phi \nabla_e \phi } \geq 0 \, , 
\end{eqnarray} 
while the strong 
energy condition ($\left( T_{ab} - T g_{ab}/2 \right) t^a t^b \geq 0$ for 
all timelike vectors $t^a$ \cite{Waldbook}) is 
\begin{eqnarray}
 \left( T_{ab}^{(\phi)} - \frac{1}{2}\, T^{(\phi)} g_{ab} \right) u^a u^b 
&\!\!=\!\!& \frac{1}{2} \left( \rho^{(\phi)} + 3P^{(\phi)} \right) 
\nonumber\\
&&\nonumber\\
&\!\!=\!\!& - \frac{\omega}{\phi^2} \, \nabla^e
\phi \nabla_e \phi - \frac{V}{2\phi} + \frac{1}{\phi} \left[ -\frac{1}{2} 
\square \phi - \frac{ \nabla^a \phi \nabla^b \phi \nabla_a \nabla_b \phi}{ 
\nabla^e \phi \nabla_e \phi } \right] \ge 0 \,.  \nonumber\\
& &
\end{eqnarray}
\section{Eckart's thermodynamics for scalar-tensor gravity} \label{sec:3} 
\setcounter{equation}{0}

In Eckart's thermodynamics \cite{Eckart40} (see also 
Refs.~\cite{Maartens:1996vi, RuggeriSugiyama21} for a pedagogical 
exposition and~\cite{Andersson:2006nr} for relativistic fluids in general), 
the dissipative quantities ({\em i.e.}, viscous pressure $P_\text{vis}$, 
heat current density $q^c$, and anisotropic stresses $\pi_{ab}$) are 
related to the expansion $\theta$, temperature ${\cal T}$, and shear tensor 
$\sigma_{ab}$ by the constitutive equations \cite{Eckart40} 
\begin{eqnarray} P_\text{vis} &=& -\zeta \, \theta \,,\\
&&\nonumber\\
q_a &=& -{\cal K} \left( h_{ab} \nabla^b {\cal T} + {\cal T} \dot{u}_a 
\right) \,, \label{Eckart}\\
&&\nonumber\\
\pi_{ab} &=& - 2\eta \, \sigma_{ab} \,, \end{eqnarray} where $\zeta$ is the 
bulk viscosity, ${\cal K}$ is the thermal conductivity, and $\eta$ is the 
shear viscosity.

The comparison of Eqs.~(\ref{eq:q}) and (\ref{acceleration}) yields \be 
q_a^{(\phi)} = -\frac{ \sqrt{-\nabla^c \phi \nabla_c \phi}}{ 8 \pi \phi} \, 
\dot{u}_a \,.\label{q-a} \ee Not only is this vector purely spatial, which 
we already know, but it is proportional to the 4-acceleration of the 
effective fluid ``elements''.

\subsection{The temperature of scalar-tensor gravity}

Equation~(\ref{q-a}), already obtained in 
Ref.~\cite{Faraoni:2018qdr}, is interpreted in the context of Eckart's 
first order (non-causal)  thermodynamics \cite{Eckart40}, in which the 
constitutive relation for the heat flux density (generalized Fourier law) 
is given by Eq.~(\ref{Eckart}). Then, Eq.~(\ref{q-a}) allows us to make the 
identifications \be {\cal K T} = \frac{ \sqrt{-\nabla^c \phi 
\nabla_c\phi}}{8\pi \phi} \,,\label{temperature} \ee and \be 
\label{eq:gradient} h_{ab} \nabla^b {\cal T} = 0 \, , \ee {\em i.e.}, the 
spatial temperature gradient vanishes identically and the heat flow arises 
solely from the inertia of energy (a possible contribution to the heat flux 
first discovered by Eckart \cite{Eckart40}).

Eq. \eqref{temperature} can then be used to identify the Eckart temperature 
of the $\phi$-fluid \cite{Faraoni:2018qdr}, which we dub ``temperature of 
scalar-tensor gravity''. It is reassuring that ${\cal K T}$ is positive 
definite, which could not have been taken for granted in a formal 
identification of quantities which, {\em a priori}, could generate either 
sign. Furthermore, ${\cal K T}$ formally vanishes when $\phi=$~const., 
which corresponds to GR and to the disappearance of the $\phi$-fluid.

\subsection{The effective viscosity of scalar-tensor gravity}

The structure of the effective imperfect fluid of the Brans-Dicke-like 
scalar field $\phi$ was chosen so that it does not contain a bulk viscosity 
term. Therefore, the bulk viscosity $\zeta=0$, but there is shear 
viscosity. This choice of splitting between the isotropic pressure term and 
the viscous one does not affect the generalized Fourier law and the 
definition of temperature of the $\phi$-fluid.

In order to calculate the shear viscosity $\eta$, it is sufficient to 
compare the anisotropic stress tensor~(\ref{piab-phi}) with the shear 
tensor~(\ref{sheartensor}). Rather surprisingly, these two expressions 
match term to term and are proportional to each other. Eckart's 
constitutive relation $\pi_{ab} =-2\eta \, \sigma_{ab}$ is satisfied if \be 
\eta= - \frac{ \sqrt{-\nabla^c \phi \nabla_c \phi}}{16\pi \phi} \ee or, 
using the expression~(\ref{temperature}), \be \label{eq:eta} 
\eta=-\frac{{\cal K T}}{2} \,. \ee The effective shear viscosity of 
scalar-tensor gravity is negative and vanishes at ${\cal K T}=0$, the GR 
case corresponding to equilibrium, to $\phi=$~const., and to the 
disappearance of the effective $\phi$-fluid. Negative viscosities are 
common in fluid mechanics, including jet streams, ocean currents, liquid 
crystals, and many other phenomena. They are turbulent (as opposed to 
molecular) viscosities and appear in systems into which energy is fed from 
the outside (see, {\em e.g.}, 
\cite{negviscosity-1,negviscosity-2,negviscosity-3,negviscosity-4}). In our 
case, there is no obvious turbulent interpretation (indeed, the 
$\phi$-fluid is irrotational) but, as a thermodynamical system, the 
$\phi$-fluid is not isolated. In the action~(\ref{STaction}), $\phi $ 
couples explicitly to gravity through the
 term $\phi {\cal R}$ mixing scalar and tensor degrees of freedom and 
exchanges energy with its thermodynamical ``surroundings''.

In the different context of spacetime thermodynamics, 
Ref.~\cite{Chirco:2010sw} identified shear viscosity as the source of 
dissipation and pointed to the absence of bulk viscosity in $f({\cal R})$ 
gravity, contrary to the previous proposal of Ref.~\cite{Eling:2006aw}. The 
results of \cite{Chirco:2010sw} are echoed in our different approach, in 
the wider class of scalar-tensor theories.

\subsection{Entropy generation and the second law}

In Eckart's thermodynamics, the entropy due to the heat flux is 
$R^a=q^a/{\cal T}$ which, in the comoving frame, has components 
\cite{Eckart40,Andersson:2006nr,Maartens:1996vi} \be R_{\mu}= 
\frac{q_{\mu}}{ {\cal T}} = \left( 0, \frac{ \vec{q}}{ {\cal T}} \right) 
\,. \ee The entropy current density in a fluid with particle density $n$ 
and entropy density $s$ is \be s^a = sn u^a +R^a = snu^a +\frac{q^a}{ {\cal 
T}} \,, \ee where $R^a$ describes entropy generation due to dissipative 
processes. While, for an isolated system, entropy is conserved in a 
non-dissipative fluid ($\nabla_c s^c=0$), in a dissipative one it is 
$\nabla_c s^c >0$ due to the entropy generation described by the vector 
$R^a$.
 
The entropy density is obtained from the first law \be {\cal T} dS= dU +PdV 
\,, \ee which yields \be s\equiv \frac{dS}{dV}= \frac{\rho+P}{{\cal T}} \ee 
assuming a closed (yet, not isolated) system.

The expressions~(\ref{effdensity}), (\ref{effpressure}), and 
(\ref{temperature}) of the effective density, pressure, and temperature of 
the $\phi$-fluid then give \begin{eqnarray}
&& s = \frac{ K}{ \sqrt{-\nabla^e \phi \nabla_e \phi}}
 \left[ -\frac{\omega}{\phi} \, \nabla^e\phi \nabla_e \phi 
+\frac{\Box\phi}{3} -\frac{4}{3} \, \frac{ \nabla^a\phi \nabla^b \phi 
\nabla_a\nabla_b \phi}{ \nabla^e\phi \nabla_e \phi} \right]\,.\nonumber\\
&&
\end{eqnarray}

In a fluid in which the particle number is conserved, $\nabla_a n^a=0$ 
(where $n^a = nu^a$ is the particle current density), one has 
\cite{Eckart40, Andersson:2006nr,Maartens:1996vi} \be \nabla_c s^c = \frac{ 
P_\text{vis}^2}{\zeta {\cal T}} +\frac{ q_c q^c}{ {\cal K}{\cal
 T}^2} + \frac{\pi_{ab} \pi^{ab}}{2\eta {\cal T}} \,, \ee where the bulk 
viscosity term (the first term on the right hand side) is absent for the 
effective $\phi$-fluid. Using the fact that $q^a=- {\cal KT} \dot{u}^a$, 
Eckart's entropy generation term is \be R^a = -{\cal K}\dot{u}^a \,. 
\label{eq:R} \ee Equations~(\ref{q-a}) and (\ref{piab-phi}) are then used 
to compute $\nabla_c s^c$, obtaining \begin{eqnarray} \frac{q_c q^c}{ {\cal 
K} {\cal T}^2} &=& {\cal K} \dot{u}_c \dot{u}^c = \frac{{\cal K}}{ 
(-\nabla^e \phi \nabla_e \phi)^3 } \left[ -\nabla^e \phi \nabla_e \phi 
\nabla_b \phi \nabla^d \phi \nabla^b \nabla^a \phi \nabla_d \nabla_a \phi + 
\left( \nabla^a \phi \nabla^b \phi \nabla_a \nabla_b \phi \right)^2 \right] 
\,, \nonumber\\
& & \\
&&\nonumber\\
\pi_{ab} \pi^{ab} &=& 8\eta^2 \sigma^2 =2{\cal K}^2 {\cal T}^2 \sigma^2 
\nonumber\\
&&\nonumber\\
&=& \frac{ \left(-\nabla^e\phi \nabla_e\phi\right)^{-2} }{32\pi^2 \phi^2}
\left\{\frac{1}{2} \left(-\nabla^e\phi \nabla_e\phi \right)^2 \left[ 
\nabla^a \nabla^b \phi \nabla_a \nabla_b \phi -\frac{ \left( \Box 
\phi\right)^2}{3} \right] +\frac{1}{3} \left( \nabla^a \phi \nabla^b \phi 
\nabla_a \nabla_b \phi \right)^2 \right. \nonumber\\
&&\nonumber\\
&\, & \left. - \left( \nabla^e\phi \nabla_e \phi \right)
\left( \nabla_a \nabla_b \phi \nabla^b \nabla_c \phi-\frac{\Box\phi}{3} \, 
\nabla_a\nabla_c \phi \right) \nabla^a \phi \nabla^c \phi \right\} \,, 
\end{eqnarray} and finally \begin{eqnarray} \nabla_c s^c &=& {\cal K}\left( 
\dot{u}^a \dot{u}_a + \frac{ {\cal K} {\cal T} \sigma^2}{\eta} \right) 
\nonumber\\
&&\nonumber\\
&=& {\cal K}\left( \dot{u}^a \dot{u}_a
-\sigma_{ab}\sigma^{ab} \right) \,. \label{entropyincrease} \end{eqnarray} 
Since the second term in Eq.~(\ref{entropyincrease}) is negative, one 
cannot conclude that the entropy increases.  Indeed, if energy is injected 
into the scalar field fluid coupled to gravity, the entropy $s$ may 
actually decrease, as it happens in non-isolated systems.

A special situation (if it is possible) is the one in which the 
$\phi$-fluid is geodesic, $\dot{u}^a=0$, which always corresponds to 
$q^a=0$, $J^a={\cal T} u^a$, and decreasing entropy density, consistent 
with the fact that the entropy generation vector~(\ref{eq:R}) vanishes and 
shear contributes to decreasing $s$ because of the negative $\eta$, as 
described by Eq.~(\ref{entropyincrease}).

In modern constitutive theories, all constitutive relations are supposed to 
obey two universal principles (see, {\em e.g.}, \cite{Ruggeri}): {\em (i)} 
the objectivity principle, {\em i.e.}, independence of the observer; {\em 
(ii)} the entropy principle, according to which any solution of a system of 
constitutive equations satisfies an additional entropy balance law with a 
non-negative entropy production. Therefore, the result in 
Eq.~\eqref{entropyincrease} suggests potential violations of the latter and 
a consequent problem in fitting this analogy with the $\phi$-fluid into the 
standard framework of constitutive theory. However, it is important to 
point out that the $\phi$-fluid is hardly a real fluid~--~indeed, its 
energy density $\rho ^{(\phi)}$ can be negative~--~and the correspondence 
with Eckart's theory comes as a mere comparison of kinetic and kinematic 
quantities characterizing the $\phi$-fluid. Furthermore, as already 
stressed, this exotic fluid is {\em not} isolated since $\phi$ couples 
directly to the gravity sector, which necessarily affects the entropy 
balance. Hence, the analogy between properties of the $\phi$-fluid and 
Eckart's thermodynamics holds provided that one keeps these {\em caveats} 
in mind.
\subsection{Possible physical interpretations} \label{sec:interp}
Since the thermodynamic interpretation of this analogy depends heavily on 
the specific choice of the solution of the system~\eqref{temperature} and 
\eqref{eq:gradient}, one can propose different formulations of the proposed 
approach without altering the physical scenario at hand. Here we discuss 
two simple possibilities.

Isolating the temperature in Eq.~\eqref{temperature} and then inserting the 
corresponding expression into Eq.~\eqref{eq:gradient} reduces the latter to 
\be \label{eq:lamerda} h_{ab} \, \nabla^b \ln \left( \frac{\sqrt{-\nabla^c 
\phi \nabla_c \phi}}{{\cal K}} \right) = 0 \, . \ee

One of the simplest solutions of this equation is \be \mathcal{K} = C \, 
\sqrt{-\nabla^c \phi \nabla_c \phi} \, , \ee with $C$ a positive constant. 
Setting, for instance, $C=1/8 \pi$ yields \be \mathcal{T} = \frac{1}{\phi} 
= G_{\rm eff} \ee and \be \mathcal{K} = \frac{\sqrt{-\nabla^c \phi \nabla_c 
\phi}}{8 \pi} \, . \ee This simple solution of Eq.~\eqref{eq:lamerda} sets 
the stage for a curious interpretation of the thermodynamic properties of 
the $\phi$-fluid. Specifically, the temperature measures the effective 
strength of the gravitational interaction while the thermal conductivity 
keeps track of the norm of $\nabla_a\phi$, and therefore of the variability 
of $\phi$. Now, if one looks at the GR limit of the theory 
$\phi=$~const., $\mathcal{T}$ should reduce to Newton's constant whereas 
$\mathcal{K}$ vanishes. In other words, the GR limit of scalar-tensor 
gravity corresponds to the ``perfect insulator'' limit for the 
$\phi$-fluid.

Alternatively, if one considers a more involved solution of 
Eq.~\eqref{eq:lamerda} such that $\mathcal{K} \neq 0$, one finds an 
alternative description for the GR limit of scalar-tensor gravity: this 
limit corresponds to ${\cal T} \to 0$, {\em i.e.}, GR corresponds to the 
absolute zero (minimum possible temperature) of the $\phi$-fluid. However, 
finding an explicit general expression for ${\cal K} = {\cal K} (\phi, 
\nabla \phi)$ becomes much more involved.
%
%
\section{The approach to the GR equilibrium state} \label{sec:4} 
\setcounter{equation}{0}

{\em A posteriori}, the expression of ${\cal K}{\cal T}$ can be 
differentiated to obtain an evolution equation for this quantity. Although 
this may seem redundant since we already know the solution of this 
equation, the latter plays the role of an effective heat equation for the 
$\phi$-fluid and is useful to understand better when, and how, the GR 
equilibrium state is approached.

The differentiation of Eq.~(\ref{temperature}) yields \begin{eqnarray} 
\frac{d\left( {\cal K}{\cal T}\right)}{d\tau} & \equiv & u^c \nabla_c 
\left( {\cal K}{\cal T}\right) = -\frac{ \sqrt{-\nabla^e\phi 
\nabla_e\phi}}{8\pi \phi} \, \frac{1}{\phi} \, \frac{ 
\nabla^c\phi\nabla_c\phi}{ \sqrt{-\nabla^e\phi\nabla_e\phi} } 
-\frac{u^c}{8\pi\phi} \, \frac{\nabla^e \phi \nabla_c\nabla_e\phi}{ 
\sqrt{-\nabla^e\phi \nabla_e\phi}} \nonumber\\
&&\nonumber\\
&=& \frac{{\cal K}{\cal T}}{\phi} \, \sqrt{-\nabla^e\phi \nabla_e\phi}
-{\cal K}{\cal T} \left( \theta- \frac{\Box\phi}{ \sqrt{-\nabla^e\phi 
\nabla_e\phi}} \right) \,. \end{eqnarray} Using again 
Eq.~(\ref{temperature}), one has \be \frac{d\left( {\cal K}{\cal 
T}\right)}{d\tau} = 8\pi \left( {\cal K}{\cal T}\right)^2 -\theta\, {\cal 
K}{\cal T} +\frac{\Box\phi}{ 8\pi \phi } \,. \label{approach} \ee It is not 
easy to interpret this equation in general since $\Box \phi $ does not have 
definite sign and the expansion $\theta$ depends in a rather complicated 
way from $\phi$ and its derivatives, however we can restrict to simple 
situations in order to gain insight into the approach to equilibrium. 
Consider the case of electrovacuum, $\omega=$~const., and $V(\phi) \equiv 
0$; then Eq.~(\ref{BDfe2}) yields $\Box\phi=0$. If $\theta<0$, then $d( 
{\cal K}{\cal T} )/d\tau > 8\pi ({\cal K}{\cal T})^2$ and ${\cal K}{\cal 
T}$ grows superexponentially, exploding in a finite time $\tau$ and 
diverging away from the GR equilibrium state. Therefore, one expects that 
near spacetime singularities, where worldlines of the $\phi$ field converge 
and $\theta<0$, the deviations of scalar-tensor gravity from GR will be 
extreme (this idea is tested in Sec.~\ref{sec:5}).

When $\theta>0$, it is possible that the negative term $-\theta {\cal 
K}{\cal T} $ in the right hand side of Eq.~(\ref{approach}) dominates over 
the positive term $8\pi ({\cal K}{\cal T})^2$, and that the solution ${\cal 
K}{\cal T}$ asymptotes to zero, approaching the GR equilibrium state. 
However, if ${\cal K}{\cal T}$ is large, the positive term will dominate 
the right hand side and lead the solution away from GR, the term linear in 
${\cal T}$ becoming negligible. Therefore, the approach to the GR 
equilibrium state is not granted and should not be expected all the time. 
In the next section, we report analytical solutions of scalar-tensor 
gravity where this diffusion occurs and others where it does not.
\section{Examples: analytical solutions of scalar-tensor gravity} 
\label{sec:5} \setcounter{equation}{0}

Here we examine certain exact solutions of scalar-tensor gravity to test 
the thermodynamical formalism of the effective $\phi$-fluid presented in 
the previous sections.

\subsection{The special case of FLRW universes}
In FLRW universes, the purely spatial heat flux density $q^c$ and the 
anisotropic stresses $\pi_{ab}$ vanish identically as a consequence of 
spatial homogeneity and isotropy, and the $\phi$-fluid reduces to a perfect 
fluid. This result is true also in Lovelock and $f({\cal R}, {\cal G})$ 
theories, where ${\cal G} \equiv {\cal R}^2-4{\cal R}_{ab} {\cal R}^{ab}+ 
{\cal R}_{abcd} {\cal R}^{abcd}$ is the Gauss-Bonnet integrand, in 
arbitrary dimension \cite{Gurses:2020kpv}, and presumably in other theories 
as well. Due to the absence of a spatial heat flux, we provisionally assign 
zero temperature ${\cal T}$ to FLRW spaces. This may not be the end of the 
story though: one can consider the possibility that, in FLRW universes, the 
heat flux becomes a timelike vector aligned with the four-velocity $u^c $ 
of comoving observers. This situation preserves the spatial homogeneity and 
isotropy of FLRW space. Then, Eckart's equation (\ref{q-a}) could only hold 
if the four-acceleration of the fluid is timelike. Indeed, this is 
precisely what happens in FLRW spaces sourced by a perfect fluid. The 
acceleration vanishes for a pressure-free dust while, for any other perfect 
fluid, there is a pressure gradient $\nabla_a P$ which is timelike, to 
preserve spatial isotropy, and points along the tangent $u_a$ to the fluid 
trajectory. The latter is not geodesic because of the pressure gradient 
$\nabla_a P$, but it is quasi-geodesic \cite{Faraoni:2020ejh}: the curve, 
as a set of points, coincides with the geodesic but the proper time is not 
an affine parameter along it. A quasi-geodesic coincides with a 
non-affinely parametrized geodesic when the proper time of the fluid (which 
is, in general, the cosmic comoving time but differs from the proper time 
of a freely falling observer), is used as a parameter 
\cite{Faraoni:2020ejh}. This is one of the few situations (others are 
listed in Ref.~\cite{Faraoni:2020ejh}) in which the four-acceleration 
$\dot{u}^a$ of a particle is not orthogonal, indeed, it is parallel, to its 
four-velocity $u^a$. Dealing with a timelike heat current density requires 
an extension of the formalism of \cite{Pimentel89,Faraoni:2018qdr} that 
explicitly requires the gradient $\nabla_c \phi$ to be timelike. While this 
extension may be possible, it goes beyond the purpose of the current 
manuscript and the peculiar situation of FLRW spaces with respect to 
Eckart's thermodynamics of scalar-tensor gravity will be discussed in 
detail in a separate publication.

\subsection{A Brans-Dicke solution with a central singularity}
We now search for an example of a solution of the scalar-tensor field 
equations with a spacetime singularity, to test whether ${\cal T} 
\rightarrow +\infty $ there and whether the deviation from the 
corresponding GR solution is significant. The general vacuum, static, 
spherically symmetric and asymptotically flat solution of the Brans-Dicke 
field equations with $V(\phi)=0$ that is not a black hole is known and has 
a central singularity (for appropriate parameter values) 
\cite{BronnikovAPP, Faraoni:2018mes}, but the corresponding scalar field 
gradient $\nabla_a \phi$ is spacelike. We look instead for dynamical 
solutions with timelike $\nabla_c \phi$, and we disregard FLRW universes in 
which the imperfect $\phi$-fluid quantities vanish identically. These 
criteria exclude most known analytical solutions of scalar-tensor gravity 
\cite{Faraoni:2021nhi}, but the following one, reported in 
Ref.~\cite{Banijamali:2019gry} satisfies them.  This Brans-Dicke solution 
is conformal to a GR solution found in \cite{Sultana15}, which generalizes 
an old geometry found by Wyman \cite{Wyman81} by including a positive 
cosmological constant $\Lambda$. The scalar field potential in the Jordan 
frame is \begin{eqnarray} V(\phi)=\frac{m^{2}\phi^{2}}{2} \end{eqnarray} 
where $ m^{2}= 2\Lambda/\kappa>0 $ and $\kappa=8\pi G$. The line element 
and Brans-Dicke scalar read 
\begin{eqnarray} 
ds^{2} &=& -\kappa 
r^{2}d\tau^{2}+\Big(1-\frac{\tau}{\tau_{\ast}}\Big)^{2} 
\Big(\frac{2dr^{2}}{1-\frac{2\Lambda r^{2}}{3}}+r^{2}d\Omega^{2}_{(2)}\Big) 
\,, \nonumber\\
&& \label{new1}\\
\phi(\tau) &=& \frac{\phi_{\ast}}{\Big(1-\frac{\tau}{\tau_{\ast}} 
\Big)^{2}} \,, \label{new2} 
\end{eqnarray} 
where $d\Omega_{(2)}^2 
\equiv d\vartheta^2 +\sin^2 \vartheta \, d\varphi^2$ is the line element on 
the unit 2-sphere, $\omega$ and $\Lambda$ are parameters of the theory, 
while $\phi_{\ast}$ arises from an initial condition. The Ricci scalar is 
\begin{eqnarray} 
\mathcal{R} &=& \frac{\omega}{\phi^{2}}\nabla^{c}\phi 
\nabla_{c}\phi+\frac{3\Box\phi}{\phi}+\frac{2V}{\phi} \nonumber\\
&&\nonumber\\
& = & \frac{1}{\kappa\Big(1-\frac{\tau}{\tau_{\ast}}\Big)^{2}}
\Big(2\Lambda\phi_{\ast}-\frac{4\omega}{\tau_{\ast}^{2}r^{2}}\Big) \,. 
\label{Ricci} 
\end{eqnarray} 
For any value of $\omega$, ${\cal R}$ diverges 
as $\tau\rightarrow\tau_{\ast}^{-}$, corresponding to a Big Crunch 
singularity, where also $\phi$ diverges.

If $\omega\neq0$, $\mathcal{R}$ diverges also when $r\rightarrow0^{+}$ (see 
below for the case $\omega=0$). The areal radius \begin{eqnarray} 
R(\tau,r)=\Big(1-\frac{\tau}{\tau_{\ast}}\Big)r \,, \end{eqnarray} vanishes 
as $r\rightarrow 0$ and there is a central singularity if $\omega\neq0$.

The slices of constant time are finite with $0\leq r \leq r_*$, where \be 
r_* = \sqrt{\frac{3}{2\Lambda}} 
\sqrt{1-\frac{8\pi\tilde{\phi_{0}}^{2}}{|2\omega+3|\kappa}} 
=\sqrt{\frac{3}{2\Lambda}} \sqrt{1-\frac{2}{\kappa\tau_{\ast}^{2}}} \ee 
(cf. Ref.~\cite{Faraoni:2021vpn}). We have, therefore, a naked central 
singularity embedded in a finite inhomogeneous universe created by 
$\Lambda$ and $\phi$, which ends at the finite future $\tau_{\ast}$.

The case $\omega=0$ was studied in Ref.~\cite{Banijamali:2019gry}. The 
curvature invariant \begin{eqnarray} \mathcal{R}_{ab}\mathcal{R}^{ab}
&=&\frac{1}{\phi^{2}}\Big(\nabla_{a}\nabla_{b}\phi\nabla^{a}\nabla^{b}\phi
+\frac{\Lambda^{2}}{\kappa^{2}}\Big) \nonumber\\
&&\nonumber\\
&=& \frac{1}{\tau_{\ast}^{4}\kappa
r^{4}\Big(1-\frac{\tau}{\tau_{\ast}}\Big)^{4}} 
\Big(\frac{9}{\kappa\tau_{\ast}^{2}}-4+\frac{8\Lambda r^{2}}{3}\Big) 
\nonumber\\
&&\nonumber\\
&\, & +\frac{\Lambda^{2}}{\kappa^{2}\phi_{\ast}^{2}}\Big(1
-\frac{\tau}{\tau_{\ast}}\Big)^{4} \label{boh} \end{eqnarray} diverges as 
$r\rightarrow 0^{+}$ (or as the areal radius $R\rightarrow 0^{+}$), 
therefore the naked central singularity persists for $\omega=0$.

This Brans-Dicke solution is also a solution of purely quadratic $f({\cal 
R})$ gravity \cite{Banijamali:2019gry} \be f({\cal R})=\frac{\kappa {\cal 
R}^2}{4\Lambda} \,. \ee This theory exhibits a restricted scale-invariance 
and does not admit a Newtonian limit \cite{PechlanerSexl66}, however it 
approximates the Starobinski model $f({\cal R})= {\cal R} +\alpha {\cal 
R}^2$ of inflation \cite{Starobinsky}, which fairs very well in the light 
of current cosmological observations \cite{Starob-obs1,Starob-obs2}. For 
this solution, the gradient $\nabla_c \phi$ is timelike and 
Eq.~(\ref{temperature}) reduces to \be {\cal K}{\cal T} = \frac{2}{ \left( 
8\pi \right)^{3/2}
 \, r \left(\tau_*-\tau \right) } \, . \ee Thus, ${\cal K}{\cal T}$ 
diverges as $r\rightarrow 0$ at the central singularity, and also at the 
Big Crunch singularity $\tau \rightarrow {\tau_*}^-$.  The corresponding GR 
solution with a positive $\Lambda$ and $\phi=$~const. is de Sitter space, 
which has no spacetime singularities. This example illustrates the previous 
assertion that, where ${\cal K}{\cal T}$ diverges, the deviation of 
scalar-tensor gravity from GR becomes extreme.

\subsection{Scalar-tensor black holes}
Vacuum, asymptotically flat, stationary black holes coincide with those of 
GR, according to a host of no-hair theorems originating from an early 
theorem by Hawking \cite{Hawking:1972qk}. This result was extended to more 
general scalar-tensor theories with varying Brans-Dicke coupling 
$\omega(\phi)$ and a potential $V(\phi)$ 
\cite{Bekenstein96,Sotiriou:2011dz}, provided that the latter has a minimum 
in which the scalar $\phi$ can lodge in stable equilibrium. The known 
exceptions to these no-hair theorems are maverick solutions in which the 
scalar $\phi$ diverges on the horizon, {\em e.g.}, in the 
Bronnikov-Bocharova-Melnikov-Bekenstein extremal black hole for a 
conformally coupled scalar \cite{BBMB-1,BBMB-2}. The proof of Hawking's 
theorem consists of showing that $\phi$ must be constant outside the 
horizon, and then gravity reduces to Einstein gravity in that region. 
Adopting the second interpretation of the thermodynamic analogy discussed 
in Sec.~\ref{sec:interp}, one has that this occurrence corresponds to zero 
``theory temperature'' ${\cal T}$ of the $\phi$-fluid. The physical 
interpretation is that, when they form, scalar-tensor black holes freeze 
the extra dynamical degree of freedom $\phi$ outside their horizons. 
However, the singularity inside the horizon becomes ``hot'' and deviates 
from GR, an idea that we intend to explore in the future. Then the no-hair 
theorems state that, outside the horizon, GR black holes are the states of 
``lowest temperature'' in the space of scalar-tensor black holes.

\subsection{Thermodynamics of stealth solutions}

We now examine the thermodynamics of stealth solutions of scalar-tensor 
gravity, {\em i.e.}, of solutions in which the geometry is the same as in 
GR with the same matter source while the scalar field is not constant, but 
does not gravitate. In other words, the defining features are that the 
effective stress-energy tensor $T_{ab}^{(\phi)}$ vanishes in that geometry 
\cite{stealth-1,stealth-2,stealth-3,stealth-4, 
stealth-5,stealth-6,stealth-7,stealth-8,stealth-9,stealth-10}, while the 
scalar field still retains a nontrivial dynamics.\\ \\ \noindent 
\textbf{Example 1}: $\omega=0$, $V(\phi) = 0$, and $\phi$ linear in time.

Consider the action in Eq.\eqref{STaction} with $\omega=0$ and $V(\phi) = 
0$. The corresponding effective stress-energy tensor reads \be 
T_{ab}^{(\phi)} = \frac{1}{\phi} \left( \nabla_a \nabla_b \phi - g_{ab} 
\Box \phi \right) \, . \ee It is then easy to see that this tensor vanishes 
if one considers the ansatz $\phi (t) = \alpha t + \beta$, with $t$ 
denoting the coordinate time of the solution of the effective Einstein 
field equations, and $\alpha, \beta$ are two positive constants. The 
linearity in time of the Brans-Dicke scalar field is actually a familiar 
feature in stealth solutions in Horndeski \cite{Babichev:2013cya} and 
beyond Horndeski theories \cite{Motohashi:2019sen, 
Babichev:2013cya,Kobayashi:2014eva, BenAchour:2018dap, Takahashi:2019oxz, 
Minamitsuji:2019shy, Babichev:2017lmw} (see also \cite{Faraoni:2021nhi}). 
Furthermore, this ansatz for the scalar field forces a restriction onto the 
energy-momentum tensor of (ordinary) matter in the theory. Indeed, from Eq. 
\eqref{BDfe2} one finds that $T^{(m)} = g^{ab} T^{(m)} _{ab} = 0$ 
throughout the spacetime manifold.

Focusing on the expressions of the general imperfect fluid representation 
of $T_{ab}^{(\phi)}$, one finds that $\rho ^{(\phi)}$, $P^{(\phi)}$, 
$q^{(\phi)}_{a}$, and $\pi^{(\phi)} _{ab}$ all vanish if the ansatz $\phi 
(t) = \alpha t + \beta$ is assumed. The same is true for the kinetic 
quantities associated with the $\phi$-fluid, {\em i.e.}, $\sigma _{ab}$, 
$\theta$, and $\omega _{ab}$. This means that Eckart's constitutive 
relations are identically satisfied by the $\phi$-fluid and one can still 
regard Eqs. \eqref{temperature} and \eqref{eq:eta} as viable definitions of 
temperature, thermal conductivity, and shear viscosity. Furthermore, it 
turns out that, for this ansatz for the Brans-Dicke scalar, \be {\cal 
K}{\cal T}= \frac{ \sqrt{ -\nabla^e\phi\nabla_e\phi}}{8\pi \phi} = 
\frac{\sqrt{-g_{00}} \, \alpha}{8 \pi ( \alpha t + \beta)} \, . \ee These 
stealth solutions correspond to non-equilibrium states of gravity since the 
scalar degree of freedom is excited and propagates, even though it does not 
gravitate. This situation can be made sense of by considering the case of 
static stealth solutions. In this case one has that $g_{00}$ is negative 
and corresponds to the norm of a timelike Killing vector field. Thus, the 
effective gravitational coupling is $G_\text{eff}\simeq 1/\phi$ and it 
evolves even though for these solutions the geometry does not, going to 
zero as $t \rightarrow +\infty$. Correspondingly, the scalar-tensor 
thermodynamic system approaches the GR state of equilibrium, or ${\cal 
K}{\cal T} \rightarrow 0$, at late times because gravity is switched off 
asymptotically.\\ \\ 
\noindent \textbf{Example~2}: Static stealth solution 
in vacuum Brans-Dicke gravity with $\omega=-1$ and $V(\phi)= V_0 \, \phi$.

The conditions $\omega=0$ and $V(\phi) = 0$ are not necessary to achieve a 
stealth solution. As an example with $\omega\neq 0, V\neq 0$, consider the 
following solution of vacuum Brans-Dicke theory found in 
\cite{Faraoni:2017ecj} (see also \cite{Faraoni:2021nhi}), which satisfies 
the field equations for $\omega=-1$ and linear potential $V(\phi)=V_0 \, 
\phi$ (with $V_0$ a positive constant).
 
The static, spherically symmetric geometry is \cite{Faraoni:2017afs, 
Faraoni:2017ecj} 
\be 
ds^2=-dt^2 +A(r)^{-\sqrt{2}} dr^2 +A(r)^{1-\sqrt{2}} 
\, r^2 d\Omega_{(2)}^2 \,,
\ee  
while the Brans-Dicke scalar field is 
\be 
\phi(t,r) = \phi_0 \, 
\mbox{e}^{2at} A(r)^{1/\sqrt{2}} 
\ee 
and $A(r) = 1-2m/r$. Here $m$ and $ a 
\neq 0 $ are parameters, while the constant $\phi_0>0 $ is related to the 
initial conditions. This geometry is a special case of the 
Campanelli-Lousto static geometry \cite{CampanelliLousto}, which is the 
form of the most general solution of the vacuum Brans-Dicke field equation 
that is static, spherically symmetric and asymptotically flat 
\cite{BronnikovAPP,Faraoni:2018mes}, but is expressed in a coordinate 
system of limited validity \cite{Faraoni:2018mes}. In general, this 
solution contains only naked singularities or wormhole throats, but not 
black holes \cite{Vanzo:2012zu,Faraoni:2018mes,Faraoni:2021nhi}.  This is 
not a stealth solution (it is straightforward to verify that, for example, 
the energy density of the effective $\phi$-fluid cannot vanish if $m\neq 
0$).  However, the limit $m\rightarrow 0$ produces a stealth solution. For 
$m=0$, the geometry reduces to the Minkowski one while $ \phi(t) =\phi_0 \, 
\mbox{e}^{2at} $ remains dynamical and does not gravitate. We have 
\be 
\nabla _c \phi = 2 a \phi \, \delta ^{0} _{\,\, c} \, , \quad \nabla^e\phi 
\nabla_e\phi= - 4 a^2 \phi^2 \, , \quad \nabla _c \nabla _d \phi = 4 a^2 
\phi \, \delta ^{0} _{\,\, c} \delta ^{0} _{\,\, d} \, , 
\ee 
from which one 
can easily infer that $\nabla _c u_d = 0$, $q ^{(\phi)} _c = 0$, and $\pi 
^{(\phi)} _{cd} = 0$. The energy density and isotropic pressure of the 
$\phi$-fluid are 
\be 
\rho ^{(\phi)} = - 2 a^2 + \frac{V_0}{2} = - P 
^{(\phi)} \, , 
\ee 
so that $T_{ab}^{(\phi)}$ vanishes identically only if 
\be 
V_0=4a^2 \,. 
\ee

The effective gravitational coupling strength reads $G_\text{eff}= 
\phi_0^{-1} \, \mbox{e}^{-2 a t} $, while 
\be 
{\cal K}{\cal T} 
=\frac{\sqrt{-\nabla^e\phi \nabla_e\phi}}{8\pi \phi} =\frac{|a|}{4\pi} 
\ee 
remains constant in time. The stealth scalar-tensor non-equilibrium state 
never approaches the GR equilibrium state $\phi=$~const.$>0$.

Note that $\nabla ^c \phi$ is parallel to the timelike Killing field $t^c = 
(\partial/\partial t)^c$. In order to have these two vector fields pointing 
in the same direction, it must be $a<0$, which implies that $G_{\rm eff} 
\sim e^{2|a|t}$ diverges exponentially at late times while ${\cal K} {\cal 
T} = |a|/4\pi$ remains constant.
\section{Conclusions and outlooks}

We have expanded and built upon the new approach to the 
thermodynamics of scalar-tensor gravity presented briefly in the previous 
Letter~\cite{Faraoni:2021lfc}, providing details. Our approach is very 
different from Jacobson's thermodynamics of spacetime 
\cite{Jacobson:1995ab, Eling:2006aw}: there, the temperature of spacetime 
is the Unruh temperature of local uniformly accelerated observers whose 
worldlines thread the fabric of spacetime, while here the temperature 
${\cal T}$ arises from Eckart's first order thermodynamics for dissipative 
fluids, which we are led to examine following the reformulation of 
scalar-tensor gravity in terms of an effective $\phi$-fluid in 
\cite{Pimentel89,Faraoni:2018qdr}. In \cite{Eling:2006aw}, it was proposed 
that, while GR is an equilibrium state of gravity, $f({\cal R})$ (which is 
a subclass of scalar-tensor)  gravity theories and, by extension, 
presumably also other modified gravity theories, constitute excited 
non-equilibrium states. Therefore, there should be a spontaneous approach 
of these excited states to the GR equilibrium state. However the equations 
ruling this approach to equilibrium, and the order parameter(s) ruling it, 
have never been identified. If alternative gravity really ``decays'' 
spontaneously to GR, it should be possible to track and model this process 
in more than one way. Indeed, modified gravity contains extra dynamical 
degrees of freedom in addition to the two massless spin two modes of GR 
contained in the metric tensor. Therefore, a theory in which these extra 
degrees of freedom are excited and propagate can rightly be called an 
``excitation'' of GR. In scalar-tensor (including metric $f({\cal R})$) 
gravity, there is only one (scalar, massive) extra degree of freedom.

In contrast with spacetime thermodynamics, our proposal is minimalist, 
using less assumptions and less fundamental ones. The new approach consists 
of the effective fluid formulation of scalar-tensor gravity plus the 
application of the constitutive relations of Eckart's first order 
thermodynamics \cite{Eckart40} to it. Eckart's non-causal thermodynamics is 
unable to describe a full relaxation process and to provide a relaxation 
time, however it gives us the effective temperature, the order parameter 
quantifying how far away a modified gravity is from GR. It provides also 
the information that bulk viscosity is zero, the explicit expression of the 
shear viscosity, a simple expression for the heat current density, as well 
the effective entropy density. One puzzling aspect is that the shear 
viscosity $\eta$ is negative, but this is not too disconcerting when one 
realizes that the thermodynamical system (the $\phi$-fluid) is not isolated 
but exchanges energy with its ``surroundings''. The most important 
consequence is that the entropy density is not necessarily forced to 
decrease, which corresponds to the fact that scalar-tensor gravity does not 
always approach the GR state. This fact opens the possibility that there 
could be other equilibrium states, {\em i.e.}, other special theories of 
gravity, at positive values of ${\cal K}{\cal T}$. The first possibility 
that comes to mind is Palatini $f({\cal R})$ gravity, in which the 
effective scalar degree of freedom is non-dynamical \cite{Sotiriou:2008rp, 
DeFelice:2010aj, Nojiri:2010wj}. However, this is not a truly new state 
since, in electrovacuo, it reduces
 to GR with a cosmological constant \cite{Sotiriou:2008rp, DeFelice:2010aj, 
Nojiri:2010wj}. The same situation occurs in cuscuton theory 
\cite{Afshordi:2006ad,Iyonaga:2018vnu} and in minimally modified gravity 
\cite{Lin:2017oow}, which could also provide non-trivial equilibrium 
states.

In spite of the limitations of Eckart's first order thermodynamics, the 
effective fluid formalism is not an analogy, but a new approach to the 
problem of describing how gravity diffuses (or not) to the GR state of 
equilibrium. A seemingly self-consistent thermodynamic theory emerges in 
this approach: ${\cal K}{\cal T}$ is positive-definite and vanishes in the 
GR equilibrium state; spherical, asymptotically flat, vacuum black holes of 
scalar-tensor gravity (except for mavericks) correspond to the ${\cal 
K}{\cal T}=0$ black holes of GR, according to the no-hair theorems; 
spacetime singularities correspond to formally infinite ${\cal K}{\cal T}$ 
and to large deviations from GR.

Several aspects of this new thermodynamics of gravity will be analyzed in 
future work, including: the special role of FLRW spaces; the idea that 
black holes have zero temperature far away from the horizon, while being 
``hot'' and deviating from GR near the singularity; and attempts to 
generalize the formalism to situations in which the gradient $\nabla^c 
\phi$ is null or spacelike. A natural generalization of the approach 
proposed here has been carried out in \cite{Giusti:2021sku} for Horndeski 
gravity. 
However, because of the much larger freedom in the choice of 
coupling functions and parameters in general Horndeski gravity, the 
intepretation of the results there is much more complicated and physical 
intuition relies on the results presented here.

In an alternative approach, one would trade temperature with chemical 
potential and assign zero temperature and entropy, but nonzero chemical 
potential, to the effective $\phi$-fluid of modified gravity. This approach 
was used in Refs.~\cite{Pujolas:2011he,Mirzagholi:2014ifa, 
LimSawickiVikman} for braided kinetic gravity (note the similarity between 
our Eq.~(\ref{q-a}) and Eq.~(3.35) of Ref.~\cite{Pujolas:2011he}). We will 
report in the future on this alternative approach for scalar-tensor 
gravity, as well as on both approaches for other theories of gravity 
alternative to GR. Eventually, the generalization from Eckart's first 
order, non-causal thermodynamics to realistic, causal extended 
thermodynamics for dissipative fluids \cite{Muller67, 
Stewart77,IsraelStewart79a, IsraelStewart79b, HisckockLindblom99, Carter91, 
MullerRuggeri98, RuggeriSugiyama21} will also have to be addressed.

\section*{Acknowledgments}

We thank Jeremy C\^ot\'e and Tommaso Ruggeri for helpful discussions  
and a referee for helpful comments. This work is supported, in part, by 
the 
Natural Sciences \& Engineering Research Council of Canada 
(Grant~2016-03803 to V.F.). A.G. is supported by the European Union's 
Horizon 2020 research and innovation programme under the Marie 
Sk\l{}odowska-Curie Actions (grant agreement No. 895648 -- CosmoDEC). 
A.M. is supported, in part, by the PRIN2017 project 
``Multiscale phenomena in Continuum Mechanics: singular limits, off-equilibrium and transitions''
(Project Number: 2017YBKNCE).
This work was also carried out in the framework of the activities of the Italian 
National Group for Mathematical Physics [Gruppo Nazionale per la Fisica 
Matematica (GNFM), Istituto Nazionale di Alta Matematica (INdAM)].

\end{document}